\documentclass[12pt]{article}
\usepackage{amssymb}
\usepackage{amsmath}
\usepackage{graphicx}
\usepackage{axodraw}
\begin{document}

\begin{titlepage}

\vspace*{1cm}

\begin{center} 
\setlength{\baselineskip}{24pt}
{\LARGE  Is the $D^+_{sJ}(2632)$ Meson a Cryptoexotic Tetraquark
Baryonium State ?}
\end{center}

\begin{center}
\vspace{2cm}
{\large B. Nicolescu}

\vspace{0.5cm} 
 Theory Group,  Laboratoire de Physique Nucl\'eaire  et des Hautes  \'Energies 
(LPNHE)\footnote{Unit\'e  de Recherche des Universit\'es 
  Paris 6 et Paris 7, Associ\'ee au CNRS.},
 CNRS and Universit\'e Pierre et Marie Curie, Paris\\ 
 {\small e-mail: \texttt{nicolesc@lpnhep.in2p3.fr }}

\vspace{1cm}
{\large J. P. B. C. de Melo\footnote{Visitor at LPNHE, CNRS and Universit\'e Pierre et Marie Curie, Paris.}}  

\vspace{0.5cm} 
 Instituto de F\' isica T\'eorica, Universidade Estadual Paulista, \\
Rua Pamplona, 145, 01405-900, S\~{a}o Paulo, SP, Brazil   \\
    {\small e-mail: \texttt{pacheco@ift.unesp.br}} 

\vspace{3cm}
\textbf{Abstract}
\end{center}
We suggest that the recently discovered charm-strange meson $D_{sJ}^+(2632)$, with unusual properties, could be a cryptoexotic tetraquark baryonium state $cd\bar d\bar s$.
We predict other four narrow states, as Regge recurrences of $D^+_{sJ}(2632)$, below the possible baryon-anti\-baryon thresholds.
\end{titlepage}
\newpage

The recently discovered charm-strange meson $D^+_{sJ}(2632)$~\cite{Evdo04} has very intriguing properties : i) it is very narrow ($\Gamma < 17$~MeV); ii) its coupling 
to the $D^+_s\eta$ channel is much stronger than its coupling to the $D^{0}~K^{+}$ channel.

The above two unusual properties indicate that the $D^+_{sJ}(2632)$ maybe a cryptoexotic tetraquark baryonium state $c d \bar d\bar s$. 
The reasons are the following:\\
\begin{itemize}
  \item [1.]
The favoured decays of a baryonium state are the baryon-antibaryon channels
(see Fig.~1):
$\Lambda_c^+$(2285)~$\bar{\Sigma}_0$(1193), 
~$\Sigma_c^0$(2455)~$\bar{\Sigma}^+$(1197) and 
$\Xi_c^0$~(2472)~$\bar\Xi^+$(1321). 
However the $D_{sJ}^+$~(2632) state has a mass well below the threshold of 
these baryon-antibaryon channels. 
Therefore its decay in these channels is forbidden. 
  \item [2.]
For a baryonium state, the meson decay channels are disfavoured as compared with the baryon-antibaryon channels. 
This last possibility being forbidden by the mass of the $D_{sJ}^+$ state, what remains are the meson channels and the hierarchy of decays is dictated by the quark content. 
A $cd\bar d\bar s$ state prefers the $D_{s}^+$~$\eta$ channel as compared with the 
$D^0K^+$ channel (see Fig.~2). 
\end{itemize}

Let us explain in some detail the selection rules governing the decay channels shown in Fig.~1 and Fig.~2. 
These selection rules were very much discussed in the $70^{\prime}$ s in 
the framework of the dual topological $1/N_C$ expansion~\cite{Roy04}. 
The crucial point here is that the gauge invariance of {\it QCD} selects the Y-shape 
of the baryon as compared with the $\Delta$-shape~\cite{Rossi} (see Fig.~3a) and 
Fig.~3b)). 
The three quarks of the baryon are joined at a junction point. 
The scattering is described by duality diagrams which are ordered in terms of the topological $1/N_C$ expansion. 
The leading terms correspond to the propagation of the junction lines (generated 
by the propagation of the junction point) from initial to final state (see Fig.~3c) and Fig. 4): this is the reason why the baryonium-baryon-antibaryon coupling is stronger than the baryonium-mesons coupling. 
This is also the reason  why the diagrams of Fig.~2a) and Fig.~2b) are suppressed as compared with the diagram of Fig.~1.

Let us make some terminological precisions. 
In the $70^{\prime}$ s one did distinguish between \textit{phaneroexotic} multiquark states and \textit{cryptoexotic} states. 
The phaneroexotic multiquark states (from the Greek \textit{phaneros}= manifest, visible, open to sight) are those states whose quark content can not be confused with the usual $q\bar q$ and $qqq$ content. 

The much celebrated $\Theta^+(1540)$ pentaquark ($uudd\bar s$) state~\cite{LEPS} is also a phaneroexotic state: its quantum numbers can not be obtained from a triquark 
combination.  

However, one must remark that evidence for phaneroexotic multiquark state was given 26 years ago from the study of ''forbidden''  forward peaks in the t-channel~\cite{Nicol78}. 
Namely in Ref.~\cite{Nicol78} we presented evidence for phaneroexotic t-channel $uu\bar d\bar d$ (I=2, S=0) exchanges in $pn \to \Delta^-\Delta^{++}$, $\pi^+n\to\pi^-\Delta^{++}$ and $\pi^-p \to \pi^{+} \Delta^{-}$, 
for $sd\bar u\bar u$ (I=3/2, S=-1) exchanges in the in $\bar pp \to \bar{Y}^{\ast +} Y^{\ast-}$, 
$\pi^- p \to K^{+} Y^{\ast-}$ and $K^{-}p \to \pi^{+} Y^{\ast-}$ and for 
$ss\bar u\bar u$, $ss\bar d\bar d$ and $ss\bar u\bar d$ (S=-2) 
exchanges in $K^-p \to K^+ \Xi^{\ast-}$. 
In the absence of baryonium exchanges, these peaks must show an energy dependence $s^{-9}-s^{-10}$ while experimentally one observes a behavior $s^{-3.5}-s^{-4.4}$.

The cryptoexotic states (from the Greek \textit{kruptos}=hidden, secret) are those  multiquark states whose quantum numbers can be also obtained from the usual $q\bar q$ or $qqq$ states. 
The discussion in the $70^{\prime}$~s was concentrated on the cryptoexotic 
tetraquark states $qq\bar q\bar q$.

In the framework of tetraquarks, a special position is played by the baryonium states: for the reason discussed above they can be very narrow, while most of the other tetraquark states are generally very broad. 
No clear evidence was given till now for a cryptoexotic baryonium state. 
The $D_{sJ}^{+}$(2632) state is possibly the first clear evidence for such a state.
It is possible that the narrow charmed meson seen in the $D_s^+\pi^0$ channel at 2317 GeV~\cite{Babar03} belongs to the same class of states \cite{Barn03}.

If our interpretation is correct, the $D_{sJ}^{+}$(2632) is the first known member of an entire family of very narrow tetraquark baryonium states.  
By assuming that all these states belong to an exchange-degenerate Regge trajectory with the universal slope $\alpha^{\prime}~\simeq~0.94\pm 0.06$~GeV$^{-2}$ (as indicated by the masses of known mesons and baryons listed in 
\textit{Review of Particle Physics}~\cite{hagi02}, 
we can compute the masses of these recurrences of the $D_{sJ}^+$(2632) state, in 
terms of the mass of the $D_{sJ}^+$ and $\alpha^{\prime}$:
\begin{equation}
m_{n}^2=\frac{n}{\alpha^{\prime}}+
m^2_{D^+_{SJ}} \text{, with } n=1,2,3 ...,
\end{equation}
We therefore predict four narrow states below the 
$\Lambda_c^+\bar\Sigma_0,\ \Sigma_c^0\bar\Sigma^+$
and $\Xi^0_c\bar\Xi^+$ thresholds:
\begin{align}
& m_1  =  2.827\pm 0.011 \text{ GeV}, & m_2 = 3.010\pm 0.022 \text{ GeV},\\ \nonumber 
& m_3  =  3.182 \pm 0.031\text{ GeV},  & m_4 = 3.345\pm 0.039 \text{ GeV}. 
\end{align}

Let us close  making some considerations on a unified terminology in view of the new 
born spectroscopy, which concerns the sector of narrow high mass resonances involving heavy quarks in their structure.
The new spectroscopy requires an unified terminology of old and new states.

In the new sector states of baryon number 1 we can define the \textit{barypolyquarks}.
\begin{equation}
q^{3+n}\bar q^n, \text{ with } n=0,1,2, ...
\end{equation}

For $n=0$, we get the usual $q^3$ baryons, which we propose to rebaptise as \textit{triquarks}. 
For $n=1$, we get the much celebrated $q^4\bar q$ \textit{pentaquarks}. 
For $n=2$ we get the $q^5\bar{q}^2$ \textit{heptaquarks}, for $n=3$ the $q^6\bar q^3$ \textit{enneaquarks}, etc.

By annihilating a barypolyquark with the corresponding antibarypolyquark  we get, \textit{via} the annihilation process of at least one $q\bar{q}$ pair, the baryon number sector $0$, 
which we propose to call \textit{mesopolyquarks}, namely the usual $q\bar q$ mesons (which we propose to rebaptise \textit{diaquarks}) and 
also the already discussed \textit{tetraquarks} $q^2\bar q^2$. 

From the requirement of keeping a minimum number of quarks and antiquarks, while still keeping a junction-antijunction pair, we define also the states 
\begin{equation}
q^{3+n}~\bar{q}^{3+n}, \text{with } n=1,2,3, ...
\end{equation}

For $n=1$ we get the $q^4\bar q^4$ \textit{octoquarks}, 
for $n=2$ the $q^5\bar q^5$ \textit{decaquarks}, 
for $n=3$ the $q^6\bar q^6$ \textit{dodecaquarks}, etc.
The octoquarks are obtained through pentaquark-antipentaquark annihilation.

The ''hexaquarks'' $q^3\bar q^3$ are absent from this list because they correspond to molecular broad states. 
The theoretical status of polyquark states other than mesons (diaquarks), baryons (triquarks), tetraquarks and pentaquarks requires intensive theoretical and 
phenomenological studies. 

\vspace{2mm}
 \textit{Note.}
After the completion of this paper, we learnt that L.~Maiani et al.
\cite{maiani} made, simultaneously with us, the assumption that
$D_{sJ}^+(2632)$ is a $cd\bar d\bar s$ baryonium state.

\vspace{2mm}
\textit{Acknowledgments.}
One of us (JPBCM) thanks the Brazilian Agency Funda\c c\~{a}o de Amparo \' a Pesquisa do Estado de S\~ao Paulo (FAPESP) for its support and the Theory Group of LPNHE Paris, where this work was performed, for its kind hospitality.

\newpage
\noindent Figure captions
\vspace{1cm}

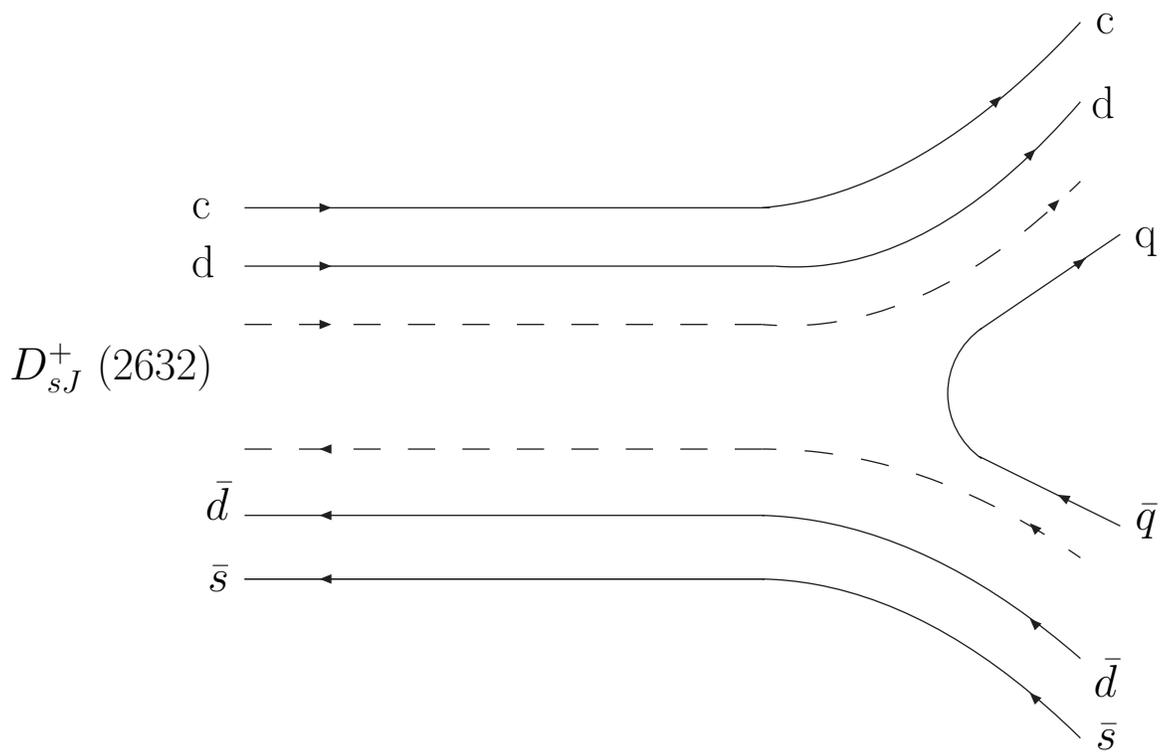
\noindent\textbf{Fig.~1.}
The decay of the $D_{sJ}^+(2632)$ meson, as a baryonium state, in the baryon-antibaryon channel.

\vspace{1cm}

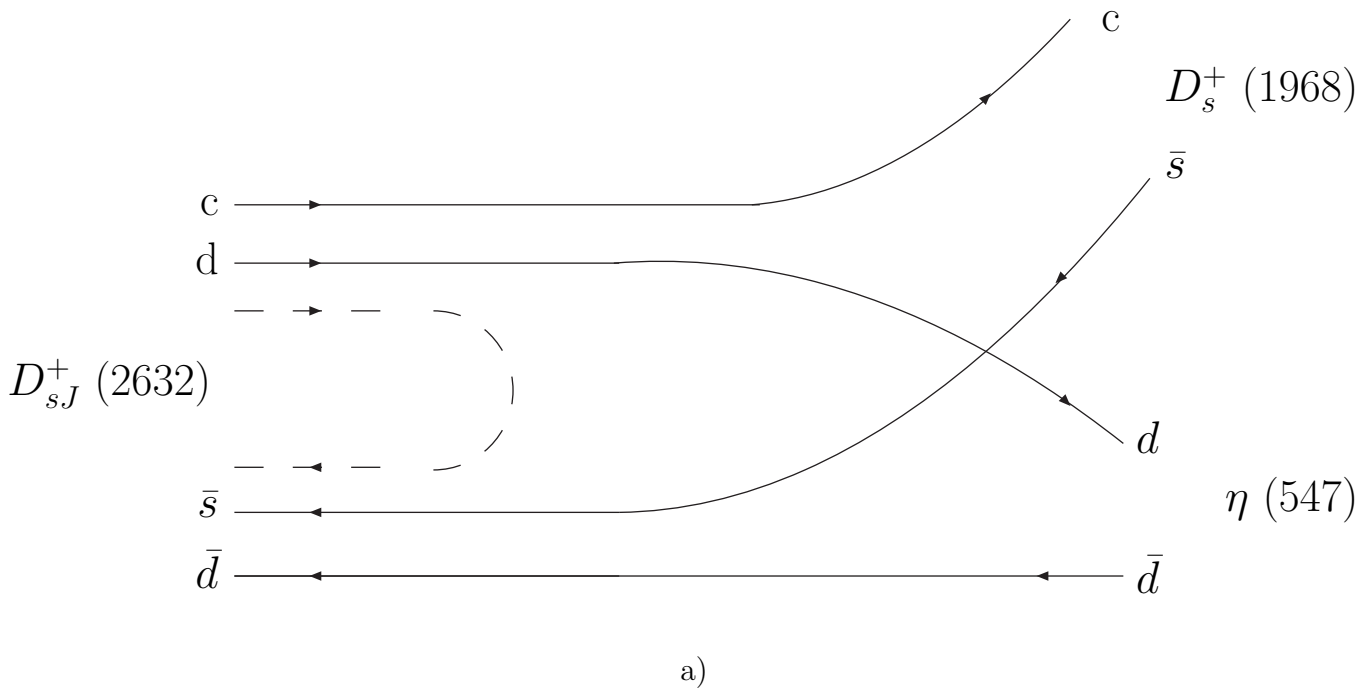
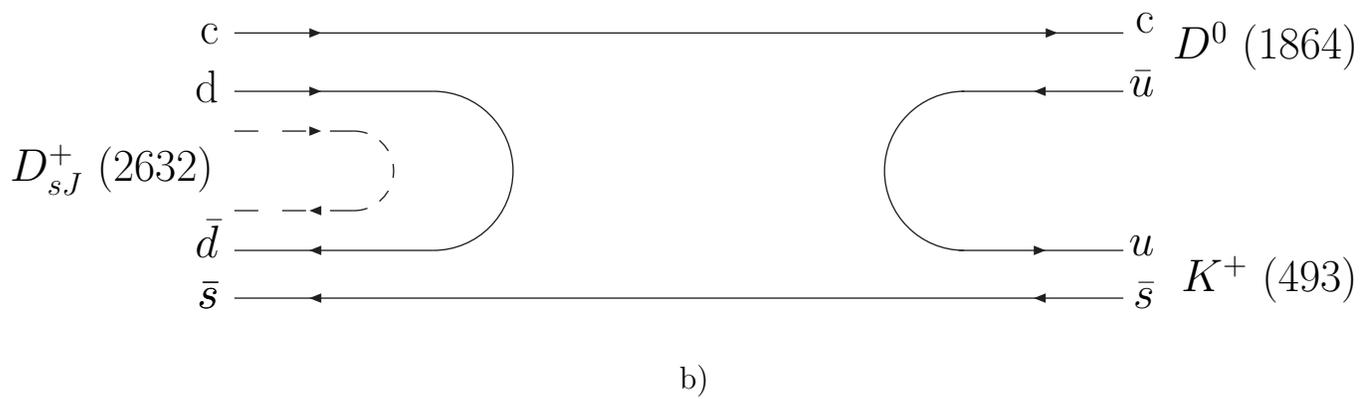
\noindent\textbf{Fig.~2.}
\begin{itemize}
  \item [a)]
The coupling of $cd\bar d\bar s$ to $D_s^+\eta$.
  \item [b)]
The coupling of $cd\bar d\bar s$ to $D^0K^+$.
\end{itemize}

\vspace{1cm}

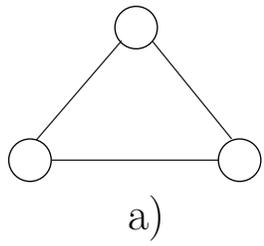
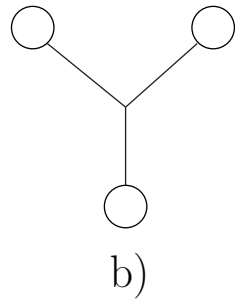
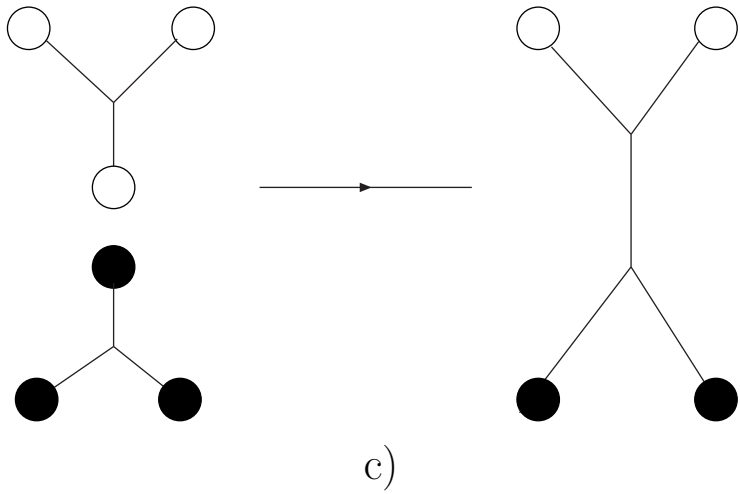
\noindent\textbf{Fig.~3.}
\begin{itemize}
  \item [a)]
Baryon as a $\Delta$-shape.
  \item [b)]
Baryon as a Y-shape.
  \item [c)]
Baryon-antibaryon pair leading to a tetraquark baryonium state.
\end{itemize}

\vspace{1cm}

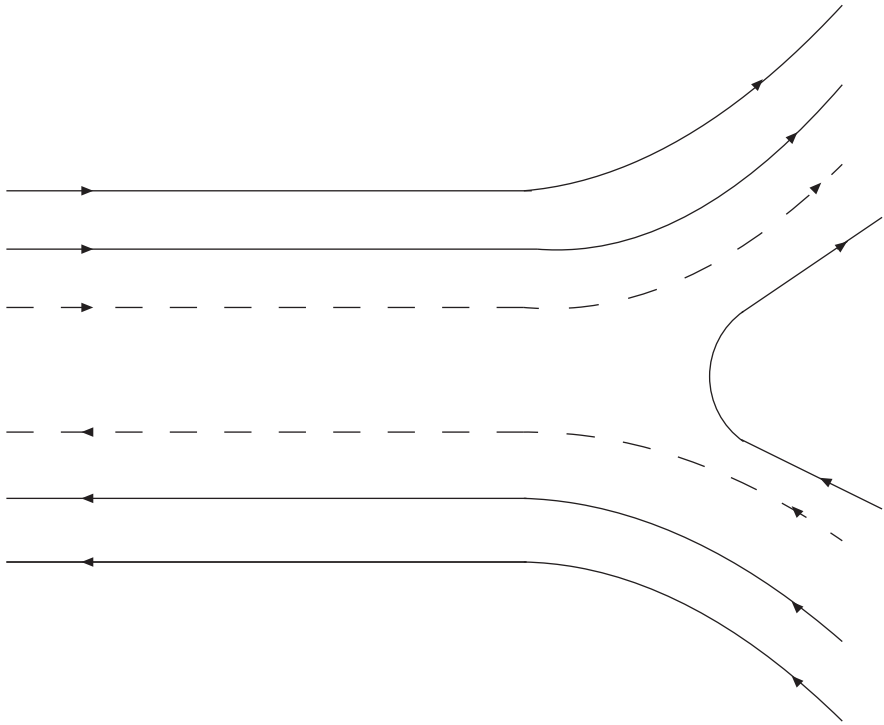
\noindent\textbf{Fig.~4.}
Baryon-antibaryon-baryonium coupling.

\newpage

\hspace*{2cm}
\begin{picture}(430,180)(0,0)
\Line(5,10)(203,10) 
\ArrowLine(35,10)(36,10)
\Curve{(200,10)(240,20)(320,80)}
\ArrowLine(288,50)(289,51)
\put(-8,6){\makebox(0,0)[br]{{\Large c}}} 
\put(-1,-16){\makebox(0,0)[br]{{\Large d }}}
\put(-7,-60){\makebox(0,0)[br]{{\Large $D_{sJ}^{+}~(2632)$}}}
\Line(5,-12)(205,-12)
\ArrowLine(35,-12)(36,-12)
\Curve{(205,-12)(220,-12)(320,50)}
\ArrowLine(301,30)(302,31)
\put(334,76){\makebox(0,0)[br]{{\Large c}}}
\put(338,44){\makebox(0,0)[br]{{\Large d} }}
\DashLine(5,-34)(200,-34){10}
\ArrowLine(35,-34)(36,-34)
\DashCurve{(200,-34)(220,-34)(320,20)}{10}
\ArrowLine(310,11)(311,12)
\Line(281.9999,-35.999)(335,0)
\CArc(300,-60)(30,125,-125)
\Line(281.888,-83.8)(335,-110)
\ArrowLine(318,-12)(322,-9)
\ArrowLine(315,-100)(312,-98.5)
\put(350,-8){\makebox(0,0)[br]{{\Large q}}}
\put(350,-115){\makebox(0,0)[br]{{\Large $\bar{q}$}}}
\Line(201,-130)(5,-130)   
\DashLine(200,-81)(5,-81){10}
\ArrowLine(36,-81)(35,-81)
\DashCurve{(200,-81)(220,-82)(320,-122)}{10}
\ArrowLine(303,-111)(302,-110)
\put(-1,-108){\makebox(0,0)[br]{\Large{$\bar{d}$}}}
\put(-1,-135){\makebox(0,0)[br]{\Large{$\bar{s}$}}}
\Line(201,-106)(5,-106)
\ArrowLine(36,-106)(35,-106)
\Curve{(200,-106)(220,-108)(320,-160)}
\ArrowLine(303,-147)(302,-146)
\put(335,-175){\makebox(0,0)[br]{{\Large $\bar{d}$}}}
\put(335,-195){\makebox(0,0)[br]{{\Large $\bar{s}$}}}
\Line(201,-130)(5,-130)
\ArrowLine(36,-130)(35,-130)
\Curve{(200,-130)(220,-132)(320,-190)}
\ArrowLine(303,-175)(302,-174)
\end{picture}
\vspace{7.5cm} 
\begin{center}
Figure 1
\end{center}

\newpage

\hspace*{1cm}
\begin{picture}(430,180)(0,0)
\Line(5,10)(203,10) 
\ArrowLine(35,10)(36,10)
\Curve{(200,10)(240,20)(320,80)}
\ArrowLine(288,50)(289,51)
\put(-4.0,-67){\makebox(0,0)[br]{{\Large $D_{sJ}^+$~(2632)}}}
\put(-1,6){\makebox(0,0)[br]{{\Large c}}}
\put(-1,-16){\makebox(0,0)[br]{{\Large d}}}
\Line(5,-12)(150,-12)
\ArrowLine(35,-12)(36,-12)
\Curve{(148,-12)(220,-18)(340,-80)}
\put(340,76){\makebox(0,0)[br]{{\Large c}}}
\put(365,20){\makebox(0,0)[br]{{\Large $\bar{s}$}}}
\put(430,45){\makebox(0,0)[br]{{\Large $D_{s}^{+}$~(1968)}}}
\ArrowLine(317,-17)(315,-19)
\ArrowLine(35,-30)(36,-30)
\DashCArc(80,-60)(30,-90,90){10}
\Line(150,-130)(5,-130)   
\DashLine(5,-30)(60,-30){10}
\DashLine(60,-89)(5,-89){10}
\ArrowLine(36,-89)(35,-89)
\Curve{(150,-106)(220,-90)(350, 20)}
\Line(150,-106)(5,-106)
\put(-1,-108){\makebox(0,0)[br]{{\Large $\bar{s}$}}}
\put(-1,-135){\makebox(0,0)[br]{{\Large $\bar{d}$}}}
\put(430,-110){\makebox(0,0)[br]{{\Large $\eta$~(547)}}}
\ArrowLine(36,-106)(35,-106)
\put(355,-137){\makebox(0,0)[br]{{\Large $\bar{d}$}}}
\put(355,-84){\makebox(0,0)[br]{{\Large $d$}}}
\ArrowLine(318,-63.8)(319,-64.8) 
\Line(340,-130)(5,-130)
\ArrowLine(36,-130)(35,-130)
\ArrowLine(310,-130)(309,-130)
\end{picture}

\vspace*{5cm}
\begin{center}
a)
\end{center}




\vspace{-4cm}

\hspace*{1cm}
\begin{picture}(430,180)(0,0)
\Line(5,10)(340,10) 
\ArrowLine(35,10)(36,10)
\ArrowLine(312,10)(314,10)
\put(-1,6){\makebox(0,0)[br]{{\Large c}}}
\put(-1,-16){\makebox(0,0)[br]{{\Large d}}}
\put(353,10){\makebox(0,0)[br]{{\Large c}}}
\put(430,-4){\makebox(0,0)[br]{{\Large$D^0$~(1864)}}}
\put(-3,-51){\makebox(0,0)[br]{{\Large$D_{sJ}^+$~(2632)}}}
\Line(5,-12)(80,-12)
\ArrowLine(35,-12)(36,-12)
\CArc(80,-42)(30,-90,90)
\Line(5,-72)(80,-72)
\ArrowLine(36,-72)(35,-72)
\DashCArc(50,-42)(15,-90,90){5}
\DashLine(5,-27)(50,-27){10}
\ArrowLine(35,-27)(36,-27)
\ArrowLine(36,-57)(35,-57)
\DashLine(5,-57)(50,-57){10}
\CArc(280,-42)(30,90,-90)
\Line(279,-12)(340,-12)
\ArrowLine(309,-12)(308,-12)
\Line(278,-72)(340,-72)
\ArrowLine(308,-72)(309,-72)
\put(353,-14){\makebox(0,0)[br]{{\Large $\bar{u}$}}}
\Line(5,-90)(340,-90)
\ArrowLine(36,-90)(35,-90)
\ArrowLine(309,-90)(308,-90)
\put(-1,-75){\makebox(0,0)[br]{{\Large$\bar{d}$}}}
\put(-1,-94){\makebox(0,0)[br]{{\Large$\bar{s}$}}}
\put(-1,-94){\makebox(0,0)[br]{{\Large$\bar{s}$}}}
\put(353,-74){\makebox(0,0)[br]{{\Large$u$}}}
\put(353,-94){\makebox(0,0)[br]{{\Large$\bar{s}$}}}
\put(435,-92){\makebox(0,0)[br]{{\Large$K^{+}$ (493) }}}
\end{picture}
\vspace*{3cm}
\begin{center}
b)\\
\vspace{0.5cm}
Figure 2
\end{center}

\newpage

\hspace*{1cm}
\begin{picture}(430,180)(0,0)
\GCirc(75,150){8}{1}
\Line(30,99)(69.5,145)
\GCirc(35,100){8}{1}
\Line(43,100)(120,100)
\GCirc(114,100){8}{1}
\Line(81,145)(111,108)
\put(86,70){\makebox(0,0)[br]{{\Large  a)}}}
\GCirc(235,150){8}{1}
\Line(240.5,144)(270,120)
\GCirc(303,150){8}{1}
\Line(297,144)(270,120)
\Line(270,120)(270,82.5)
\GCirc(270,82.5){8}{1}
\put(280,48){\makebox(0,0)[br]{{\Large b)}}}
\GCirc(33,0){8}{1}
\Line(39.5,-4.5)(65,-28)
\GCirc(95,0){8}{1}
\Line(89,-4)(65,-28)
\Line(65,-28)(65,-58)
\GCirc(65,-60){8}{1}
\GCirc(65,-90){8}{0}
\Line(65,-96)(65,-120)
\Line(65,-120)(36,-140)
\Line(65,-120)(90,-140)
\GCirc(36,-140){8}{0}
\GCirc(90,-140){8}{0}
\GCirc(225,0){8}{1}
\Line(230,-7)(260,-40)
\GCirc(292,0){8}{1}
\Line(285.5,-5)(260,-40)
\Line(260,-40)(260,-90)
\Line(260,-90)(218,-145)
\Line(260,-90)(295,-145)
\GCirc(225,-140){8}{0}
\GCirc(292,-140){8}{0}
\ArrowLine(120,-60)(200,-60)
\put(173,-175){\makebox(0,0)[br]{{\Large c)}}}
\end{picture}
\vspace{8cm}
\begin{center}
Figure 3
\end{center}

\newpage

\hspace*{1cm}
\begin{picture}(430,180)(0,0)
\Line(5,10)(203,10) 
\ArrowLine(35,10)(36,10)
\Curve{(200,10)(240,20)(320,80)}
\ArrowLine(288,50)(289,51)
\Line(5,-12)(205,-12)
\ArrowLine(35,-12)(36,-12)
\Curve{(205,-12)(220,-12)(320,50)}
\ArrowLine(301,30)(302,31)
\DashLine(5,-34)(200,-34){10}
\ArrowLine(35,-34)(36,-34)
\DashCurve{(200,-34)(220,-34)(320,20)}{10}
\ArrowLine(310,11)(311,12)
\Line(281.9999,-35.999)(335,0)
\CArc(300,-60)(30,125,-125)
\Line(281.888,-83.8)(335,-110)
\ArrowLine(318,-12)(322,-9)
\ArrowLine(315,-100)(312,-98.5)
\Line(201,-130)(5,-130)   
\DashLine(200,-81)(5,-81){10}
\ArrowLine(36,-81)(35,-81)
\DashCurve{(200,-81)(220,-82)(320,-122)}{10}
\ArrowLine(303,-111)(302,-110)
\Line(201,-106)(5,-106)
\ArrowLine(36,-106)(35,-106)
\Curve{(200,-106)(220,-108)(320,-160)}
\ArrowLine(303,-147)(302,-146)
\Line(201,-130)(5,-130)
\ArrowLine(36,-130)(35,-130)
\Curve{(200,-130)(220,-132)(320,-190)}
\ArrowLine(303,-175)(302,-174)
\end{picture}
\vspace{7cm} 
\begin{center}
Figure 4
\end{center}


\begin{thebibliography}{99}
\bibitem{Evdo04}
SELEX Collaboration: A.~V.~Evdokimov et al., hep-ex/0406045.

\bibitem{Roy04}
Proceedings of the Meeting on Exotic Resonances, Hiroshima, September 1-2, 1978, edited by I.~Endo, Y.~Sumi, S.~Wakaizumi and M.~Yonezawa;
Proceedings of the International Workshop on Baryonium and Other Unusual Hadron States, IPN Orsay, France, June 21-22, 1979, edited by B.~Nicolescu, R.~Vinh~Mau and J.-M.~Richard;
L.~Montanet, G.~C.~Rossi and G.~Veneziano, Phys. Rept. \textbf{63}, 149 (1980); for a recent review see D.~P.~Roy, J. Physics \textbf{G30}, R113 (2004).

\bibitem{Rossi} 
G.~C.~Rossi and G.~Veneziano, Nucl.~Phys.~{\bf B123}, 507 (1977).

\bibitem{LEPS} 
LEPS Collaboration: T.~Nakano et al., Phys. Rev. Lett. \textbf{91}, 012002 (2003);
CLAS Collaboration: S.~Stepanyan et al., Phys. Rev. Lett. \textbf{91}, 252001 (2003);
SAPHIR Collaboration: J.~Barth et al., hep-ex/0307083;
DIANA Collaboration: V.~V.~Barmin et al., Phys. Atom. Nucl. \textbf{66}, 1715 (2003);
Yad. Fiz. \textbf{66}, 1763 (2003).

\bibitem{Nicol78} 
B.~Nicolescu, Nucl. Phys. {\bf B134}, 495 (1978).

\bibitem{Babar03} 
BABAR Collaboration: B.~Aubert et al., Phys. Rev. Lett. \textbf{90}, 242001 (2003);
CLEO Collaboration: D.~Besson et al., AIP Conf. Proc. \textbf{698}, 497 (2004);
BELLE Collaboration, K. Abe et al., hep-ex/0307041.

\bibitem{Barn03}
T.~Barnes, F.~E.~Close and H.~J.~Lipkin, Phys. Rev. \textbf{D68}, 054006 (2003).
  
\bibitem{hagi02}
\textit{Review of Particle Physics}, K.~Hagiwara et al., \textbf{66}, 010001 (2002). 

\bibitem{maiani}
L. Maiani, F.~Piccinini, A.~D.~Polosa and V.~Riquer, hep-ph/0407025.


\end{thebibliography}
\end{document}